\documentclass[10pt,twocolumn,showpacs,showkeys,pre,aps]{revtex4-1}

\usepackage[english]{babel}
\usepackage{graphicx}
\usepackage{color}
\usepackage{graphicx}

\usepackage{array}

\usepackage{amsmath}
\usepackage{amsfonts}
\usepackage{amssymb}
\usepackage{amsthm}
\usepackage{mathrsfs}
\usepackage{dsfont}

\DeclareMathAlphabet{\mathpzc}{OT1}{pzc}{m}{it}

	\newcommand{\AsymEq}{\propto}

	\newcommand{\mr}[1]{\mathrm{#1}}			

	\newcommand{\br}[1]{\left( #1 \right)}
	\newcommand{\brr}[1]{\left[ #1 \right]}
	
	\newcommand{\of}[1]{\!\br{#1}}
	\newcommand{\off}[1]{\!\brr{#1}}
	
	\newcommand{\sbr}[1]{( #1 )}
	
	\newcommand{\sbrrr}[1]{\{ #1 \}}
	\newcommand{\sof}[1]{\!\sbr{#1}}

	\newcommand{\Sum}[2]{\sum\limits_{#1}^{#2}}

	\newcommand{\Int}[3]{\int\limits_{#1}^{#2}\mr{d}#3\,}

	\newcommand{\sInt}[3]{\int_{#1}^{#2}\mr{d}#3\,}

	\newcommand{\EA}[1]{\xpc{#1}}

	\newcommand{\xpc}[1]{\left\langle #1 \right\rangle}

	\newcommand{\sEA}[1]{\sxpc{#1}}

	\newcommand{\sxpc}[1]{\langle #1 \rangle}

	\newcommand{\Reals}{\ensuremath{\mathbb{R}} }

	\renewcommand{\d}{\mathrm{d}}

	\newcommand{\del}{\partial}
	
	\newcommand{\Landau}[1]{\mathpzc{O}\of{#1}}

		\newcommand{\Min}[2]{\min\of{#1,#2}}
		
		\newcommand{\Abs}[1]{\left\vert #1 \right\vert}
		\newcommand{\sAbs}[1]{\vert #1 \vert}

\begin{document}

	\title{Non-spectral modes and how to find them in the Ornstein-Uhlenbeck process with white $\mu$-stable noise}
	\date{\today}

	\author{F. Thiel}
	\email{thiel@posteo.de}
	\affiliation{Institut f\"ur Physik, Humboldt Universit\"at zu Berlin, Newtonstra\ss e 15, 12489 Berlin, Germany}

	\author{I.M. Sokolov}
	\email{igor.sokolov@physik.hu-berlin.de}
	\affiliation{Institut f\"ur Physik, Humboldt Universit\"at zu Berlin, Newtonstra\ss e 15, 12489 Berlin, Germany}

	\author{E.B. Postnikov}
	\email{postnicov@gmail.com}
	\affiliation{Department of Theoretical Physics, Kursk State University, Radishcheva st., 33, 305000, Kursk, Russia}

	\begin{abstract}
		We consider the Ornstein-Uhlenbeck process with a broad initial probability distribution (L\'evy distribution), which exhibits so-called non-spectral modes.
		The relaxation rate of such modes differs from those determined from the parameters of the corresponding Fokker-Plank equation.
		The first non-spectral mode is shown to govern the relaxation process and allows for estimation of the initial distribution's L\'evy index.
		A method based on continuous wavelet transformation is proposed to extract both (spectral and non-spectral) relaxation rates from a stochastic data sample.
	\end{abstract}

	\maketitle

	\section{Introduction}
		The dynamics of systems in the vicinity of a stable equilibrium subject to fluctuations presents a diffusion process governed by a Fokker-Planck equation (FPE).
		In the sufficiently general case of small displacements from a stable equilibrium and Gaussian fluctuations  (``Gaussian white noise``) that corresponds to the Ornstein-Uhlenbeck process (OUP), \cite{Uhlenbeck1930}.
		The assumption of Gaussianity can be relaxed, when stable noise of index $\mu \in (0,2]$ is used.
		The fractional Fokker-Plank equation (FFPE) takes the form  
		\begin{equation}	
				\dot{\rho}\of{x;t}
			=
				\nu \frac{\del}{\del x}\brr{ x \rho\of{x;t} }
				+ 
				K \frac{\del^\mu}{\del \Abs{x}^\mu} \rho\of{x;t}
			\label{eq:FPE}
		\end{equation}
		where $\nu$ and $K$ are the friction and diffusion coefficient, respectively. 
		$\del^\mu / (\del \Abs{x}^\mu)$ is the Riesz-Weyl fractional derivative, defined by its Fourier transform $-\Abs{k}^\mu$.
		OUPs and their generalizations are used in physics and other fields, and are especially important in finance where it is called Vasicek model, see \cite{Aalen2004,Gajda2015,Masuda2004,Vasicek1977} and references therein.

		The problem whose solution we seek is the initial value problem on the whole real axis: we are interested in the evolution and relaxation to equilibrium of the probability density function (PDF) $\rho\of{x;t}$ of the distribution of a system's state $X\of{t}$ for given initial state $\rho_\text{in}\of{x}$ and concentrate on the temporal pattern of relaxation, especially on its long-time behavior.
		The initial PDF is of course non-negative and normalized to unity.

		With help of a similarity transformation presented in \cite{Toenjes2013}, the fractional FPE can be reduced to the common OUP's FPE.
		The methods of solution of FPEs, including those based on spectral decomposition, are discussed in detail in the classical monograph \cite{Risken1989}.
		Provided the stationary distribution $\rho_\text{eq}$ exists, the initial Fokker-Planck problem can be reduced to a Schr\"odinger-like equation using another similarity transformation.
		The Fokker-Planck equation for the Ornstein-Uhlenbeck process is mapped to the quantum mechanical harmonic oscillator.
		Due to this, it is often assumed that the diffusion problem and the quantum mechanical one are isospectral.
		Therefore the relaxation of the initial distribution to the equilibrium one is expected to follow the multi-exponential pattern $\rho\of{x;t} = \sum_\lambda\rho_{\lambda}\of{x} e^{-\lambda t}$ with rates $\lambda_j = j\mu\nu$ corresponding to the equidistant spectrum of a quantum harmonic oscillator.
		Ref.~\cite{Toenjes2013} has shown that this is not always the case.
		It has been shown that rates absent in the spectrum of the Schr\"odinger operator might appear for initial conditions corresponding to probability densities which decay at infinity slow enough (as power laws).
		These rates were termed as ``non-spectral'' rates.
		More detailed description of the spectrum are given in \cite{Toenjes2014} and \cite{Janakiraman2014}.
		Ref. \cite{VanDenBroeck2015} studies relaxation of power-law initial conditions, as well, but it is mainly concerned with non-exponential relaxation due to non-linear force.
		Broad initial distributions have also been used to explain effects in scattering experiments \cite{Stanton1988} and electron mobility measurements \cite{Deveaud1991} in semi-conductors.
		The authors of \cite{Metzler2001} already considered the FPE with power-law initial data.
		Although they did discuss how the Boltzmann equilibrium is restored, they did not consider the way there to, i.e. the relaxation process itself.
		This gap is filled in this article.

		The principal goal of the present work is to show that non-spectral relaxation rates can be observed in simulations and to propose a technique to reveal ``broad'' initial conditions from the relaxation spectrum.

		Although the following manuscript considers the one-dimensional OUP, we stress that the approach is easily generalizable to higher dimensions by replacing the derivatives in Eq. \eqref{eq:FPE} with divergence and fractional Laplace operator.
		Furthermore, all considerations in Fourier domain can also be applied in the general case of infinitely divisible noise, \cite{Masuda2004}.

	\section{Variable separation without pre-selection}
		The problem we encounter is an initial value problem for the FFPE on the whole real axis.
		We note that the form of the equation guarantees that for any integrable initial condition its integral over the whole line is conserved, and moreover, that if the initial state is represented by a non-negative function, the non-negativity of solution is retained at all later times.
		Moreover, provided the stationary (equilibrium) solution exists, the solution for any initial condition converges to this one; no blowing up and no oscillations are possible.
		As we show in Appendix A, none of these properties rely on assumptions on spectral properties of the corresponding Fokker-Planck operator, and can be obtained from the equation as it is.
		These properties make explicit introduction of boundary conditions superfluous for the class of problems under consideration; introduction of boundary conditions not motivated by the physics of the problem may lead to wrong or paradoxical results.

		The fact that the discussion of non-spectral relaxation patterns was missing in the literature is connected to the standard approach of eigenvalue decomposition as discussed in Sec. 5.4 of Ref.~\cite{Risken1989}.
		This ansatz leads a multi-exponential relaxation pattern with the spectral rates $\lambda_j = j\mu\nu$ corresponding to the equidistant spectrum of a quantum harmonic oscillator.
		Here it was explicitly assumed that $\rho\of{x;t}$ decays faster than $e^{- (\nu x^2) / (4K) }$ at infinity, but the restrictions put by this assumption were not discussed.
		Those boundary conditions make an assumption on the ``correct'' parts of the operator's spectrum even before the eigenstates are found!

		The set of the eigenfunctions of the Hermitian Schr\"odinger operator is however insufficient for expansion of the growing functions, which inevitably appear for ``broad'' initial conditions with power-law tails \cite{Toenjes2014}. 
		As a result, the operator's modes correspond to the system's relaxation properties, which in turn manifest in the system's long-time behavior.
		We will therefore find the (correct) solution of the initial value problem for Eq.\eqref{eq:FPE}, and determine its relaxation pattern.

		\subsection{Relaxation in the PDF}
			As a first step, we write Eq. \eqref{eq:FPE} in Fourier space, where it becomes a first order partial differential equation:
			\begin{equation}
					\dot{\hat{\rho}}\of{k;t}
				=
					- \nu k \frac{\del}{\del k} \hat{\rho}\of{k;t}
					- K \Abs{k}^\mu \hat{\rho}
				,
				\label{eq:FPEFourier}
			\end{equation}
			where $\hat{\rho}\of{k;t} = \sInt{\Reals}{}{x} e^{i k x} \rho\of{x;t}$ is the usual Fourier transform.
			As already noted in \cite{Toenjes2013}, any FFPE for the OUP can be mapped into the equation for $\mu' = 2$ by taking $k = \kappa \Abs{\kappa}^{-1 + 2 / \mu}$.
			Eqs.\eqref{eq:FPE} and \eqref{eq:FPEFourier} are homogeneous linear differential equations, which means that their solutions satisfy the superposition principle.

			Many methods of solution of linear homogeneous equations are based on the superposition principle, which include the Green's function approach and the variable separation method (which may lead to the eigenfunction expansion).
			In these methods, the solutions to the initial value problems are build as weighted sums or integrals of candidate solutions (which we will call components) with weights chosen in such a way, that the initial condition is satisfied for $t=0$.

			The variable separation method starts from looking for components which have the form of a product $\hat{\rho}\of{k;t} = \tau_\lambda\of{t} \hat{\rho}_\lambda\of{k}$.
			Substituting this form into Eq.\eqref{eq:FPEFourier}, we find that such a solution is only possible if $\hat{\rho}_\lambda$ and $\tau_\lambda$ are eigenfunctions of the operators appearing in Eq.\eqref{eq:FPEFourier}, i.e. 
			\begin{equation}
					\frac{1}{\tau_\lambda\of{t}} \frac{\d \tau}{\d t}
				=
					-\lambda
				=
					- \nu \frac{1}{\hat{\rho}\of{k}}
					k \frac{\d}{\d k} \hat{\rho}_\lambda\of{k}
					- K \Abs{k}^\mu
				.
				\label{eq:EigProb}
			\end{equation}
			Solving the equation for the temporal part we immediately get 
			\begin{equation*}
					\tau_\lambda\of{t}
				= 
					e^{-\lambda t}
				;
			\end{equation*}
			the fact that the solution of any initial condition tends to a stationary (equilibrium) state and doesn't show oscillations implies that the relevant values of parameter $\lambda$ are real and non-negative.
			The spatial part of the equation is solved by
			\begin{equation}
					\hat{\rho}_\lambda\of{k}
				=
					C\of{k / \Abs{k} }
					\Abs{k}^{\tfrac{\lambda}{\nu}} e^{ - \tfrac{K}{\mu \nu} \Abs{k}^\mu } 
				.
				\label{eq:EigState}
			\end{equation}
			The eigenstate $\rho_\lambda$ can be identified as the $\lambda / \nu$-th fractional derivative of the stationary state $\rho_\text{eq} = \rho_0$ (which is the characteristic function of a symmetric $\mu$-stable random variable).
			The prefactor $C\sof{k / \sAbs{k}}$ determines the parity of the solution (in higher dimensions it determines the angular behavior as well).
			We will only consider the symmetric situation and put $C\sof{k/\sAbs{k}} = 1$; that means all eigenvalues except $\lambda = 0$ are degenerate.
			The solution for $\lambda = 0$ is the equilibrium, stationary solution of the initial Fokker-Planck equation.
			If the initial state can be represented as a sum 
			\begin{equation*}
					\rho_\text{in}\of{x}
				=	
					\Sum{\lambda}{} a_\lambda
					\rho_\lambda\of{x} 
				,
			\end{equation*}
			according to superposition principle, the further time evolution is known:
			\begin{equation}
						\rho\of{x;t} 
					= 
						\Sum{\lambda}{} a_\lambda
						\rho_\lambda\of{x} e^{-\lambda t}
				.
				\label{eq:SpectralDecomp}
			\end{equation}
			The mode coefficients $a_{\lambda}$ select the appropriate eigenvalues.
			The set of admissible eigenvalues $\sbrrr{\lambda} \subset \mathbb{C}$ is determined by the boundary conditions of the equation and by the initial state.
			For the Fokker-Planck equation, we have the conditions of unit normalization, positivity of the PDF, and the existence of the stationary state.
			These restrict the eigenvalues to be non-negative and real (no blow-up, no oscillations of the solution) and enforce that $\lambda = 0$ is an admissible eigenvalue.
			Then the solution will converge to the stationary state.
			Such are the natural selection rules of the spectrum of the Fokker-Planck equation.
			They can not be considered as boundary conditions for the eigenvalue problem, since they pose restrictions on the sum Eq.\eqref{eq:SpectralDecomp} and not on the eigenstates $\rho_\lambda$.
			Initial data determines the values of the mode coefficients (the weights) via Eq.\eqref{eq:SpectralDecomp}.
			The solution of the problem thus reduces to fitting the initial condition by the weighted sum (or integral) over functions from the set of the component functions.
			This was done in \cite{Janakiraman2014} by a Taylor expansion in the propagator.

			When transforming the problem to a Schr\"odinger equation, a similarity transformation must be used.
			The similarity transformation renders the normalization condition in the original problem useless.
			Additionally, one imposes the square-integrability of the Schr\"odinger-eigenfunctions, which is completely unrelated to the original problem.

			The spectral decomposition methods correspond to a pre-selection of components with the $\lambda$-values necessary to define a set of bi-orthogonal basis functions which might appear in the expansion.
			For the Fokker-Planck problem this set has to be completed by a dual set of left-eigenfunctions, which together with the (right) eigenfunctions build a bi-orthogonal system.
			Considerable simplification is therefore given by transformation of the Fokker-Planck operator to a Schr\"odinger one, which is Hermitian and possesses a orthonormal basis of eigenfunctions.
			However, as Ref.~\cite{Toenjes2014} shows, some fully legitimate initial conditions are transformed into functions growing at infinity, which are not square-integrable, and which cannot be expanded over the known eigenfunctions; the temporal decay of these functions does not follow the ``spectral'' pattern.

			The variable separation method is however applicable without any pre-selection, and the difference to a spectral approach is only pertinent to how the corresponding sums or integrals over candidate components are fitted to the initial state.
			The corresponding examples show how spectral and non-spectral relaxation patterns appear.

			Since we are not operating in the space of square-integrable functions, the series in Eq.\eqref{eq:SpectralDecomp} is not a decomposition into orthonormal base functions, but rather a formal series.
			It has to be understood in the sense of an asymptotic expansion.
			The reverse procedure is determining the mode-coefficients by asymptotically fitting the solution of Eq.\eqref{eq:FPE} to the eigenfunctions.
			In the case of the OUP, this can done via the Fourier representation of $\rho_\lambda$.
			Let us fix some upper bound for the considered exponents, $\Lambda / \nu$.
			Combining Eq.\eqref{eq:EigState} with Eq.\eqref{eq:SpectralDecomp} and expanding up to order $\Abs{k}^{\Lambda / \nu}$ bears:
			\begin{equation*}
					e^{ \tfrac{K}{\mu\nu} \Abs{k}^\mu } \hat{\rho}_\text{in}\of{k}
				=
					\Sum{\lambda < \Lambda}{} \Abs{k}^{\tfrac{\lambda}{\nu}} a_\lambda
					+ \Landau{\Abs{k}^{\frac{\Lambda}{\nu}}}
				.
			\end{equation*}
			The coefficients found from the power series expansion are the sought after values for the spectral coefficients: $a_\lambda$.
			In case the initial condition is given by a stable law with index $\alpha$, i.e. $\hat{\rho}_\text{in}\of{k} = \mathrm{exp}\of{- \Abs{\sigma k}^\alpha}$, we can expand the two exponentials and have
			\begin{equation*}
					\Sum{m,j = 0}{M} 
					\frac{ \br{\tfrac{K}{\mu\nu}\Abs{k}^\mu}^m \br{-\Abs{\sigma k}^\alpha}^j }{m! j!}
					-
					\Sum{\lambda < \Lambda}{} \Abs{k}^{\tfrac{\lambda}{\nu}} a_\lambda
				=
					\Landau{\Abs{k}^{\frac{\Lambda}{\nu}}}
				.
			\end{equation*}
			The equation can be solved by taking the mode coefficients
			\begin{equation}
					a_{m,j}
				=
					\frac{ \br{\tfrac{K}{\mu\nu}}^m \br{-\sigma^\alpha}^j }{m! j!}
				,
				\label{eq:SpectralCoeff}
			\end{equation}
			and the relaxation rates $\lambda_{m,j} =  m\mu\nu + j\alpha\nu$.

			Let us shortly discuss this result.
			The relaxation rates $m\mu\nu$ are the spectral ones, i.e. also belong to the associated Schr\"odinger operator.
			However, all other non-negative relaxation rates are admissible for a Fokker-Planck equation, too.
			Since they do not appear in the associated Schr\"odinger operator, they are considered ``non-spectral''.
			In our case, the non-spectral rates are $\alpha j \nu$, and come from the ``broad'' initial condition.
			Of course, the distinction only makes sense, when $\alpha$ is not a multiple of $\mu$.
			Note that a term proportional to $\Abs{k}^\alpha$ with $\alpha < 2$ occurs whenever the initial state lacks a finite second moment; well-behaved initial conditions can be expanded in powers of $\Abs{k}$, bearing the same spectrum as in \cite{Janakiraman2014,Janakiraman2015}.
			For an ordinary OUP, we have $\mu = 2$, and (symmetric) initial state without power-law tails always leads to spectral relaxation.
			Hence, non-spectral relaxation is a rather artificial situation in the ordinary OUP.
			For a L\'evy OUP with fractional derivative in Eq. \eqref{eq:FPE} however, non-spectral relaxation is the rule, because $\alpha$ is rarely an integer multiple of $\mu$.
			This is the case even for the very large class of well-behaved initial states, where we have $\alpha = 2$, which is not necessarily a multiple of $\mu$.
			Hence, purely spectral relaxation in L\'evy OUP can be considered ``rare''. 
			Please note, that we only considered symmetric initial conditions -- even functions $\rho_\text{in}$.
			Admitting asymmetry lifts the degeneracy of the eigenstates in Eq.\eqref{eq:EigState}, i.e. $C\of{k / \Abs{k}}$ is no longer unity.
			In the absence of broad initial state, this leads to terms proportional to $k$ in the expansion of the initial state, and consequently, the spectrum is $m\mu + j$, just as reported in \cite{Janakiraman2014}.

			The procedure described shows the exact reason for appearing of the non-spectral series: the spectrum of the Fokker-Planck operator is not defined without specifying the boundary conditions; one can say, it is trivially continuous, because the solution to any $\lambda$ does exist and is legitimate since its behavior is not restricted by any additional condition.
			The choice from the candidate solutions is done by asymptotically matching them to the initial state.
			Strictly speaking, it might happen that the choice is not unique, but in this case the results will have to be different asymptotic representations of the same solution.
			We note that the procedure does not imply the solution of the initial equation by the method of characteristics, and does not rely on the existence of the analytic form of the full solution, but just on an asymptotic expansion.

			Let us turn to the relaxation behavior of observables.

		\subsection{Relaxation in observables}
			From the spectral decomposition \eqref{eq:SpectralDecomp} we can compute the long-time behavior of any observable of the system's state.
			Let us consider some function $f\of{x}$, such that $\sInt{\Reals}{}{x} f\of{x}\rho\of{x;t}$ exists for all times.
			In particular the integrals with respect to $\rho_\text{eq}$ as well as with respect to $\rho_\text{in}$ exist.
			Let's additionally assume that $f$ is an even function of $x$.
			Then the temporal behavior of $F\of{t} = \sEA{f\of{X\of{t}}}$ is, according to Eq.\eqref{eq:SpectralDecomp}, given by:
			\begin{equation}
					F\of{t} 
				=
					\Sum{\lambda < \Lambda}{} a_\lambda e^{-\lambda t} 
					\Int{\Reals}{}{x} f\of{x} \rho_\lambda\of{x}
				=
					\Sum{\lambda < \Lambda}{} A_\lambda e^{-\lambda t} 
					+
					\Landau{e^{-\Lambda t}}
				.
				\label{eq:ObsRelax}
			\end{equation}
			In the case of spectral relaxation, the rates are eigenvalues of the associated Schr\"odinger operator.
			As we have seen, depending on the initial condition, non-spectral rates can occur as well.
			These rates occur in every (!) observable, and -- more importantly -- may dominate the complete relaxation process.

			Let us return to our example.
			If the initial state is an $\alpha$-stable distribution with $\alpha < \mu$, the first rate occurring in \eqref{eq:ObsRelax} is $\alpha\nu$ which is smaller than the first spectral rate $\mu\nu$.
			Hence the process remains a signature of its initial state, during the whole relaxation process.
			Finding the lowest relaxation rate, thus allows for testing whether the initial state was broad or not.

	\section{Inferring the initial state from the relaxation}
		We set out for finding the smallest rate $\lambda_\text{min}$ in the relaxation of some observable.
		For example we take the $\gamma$-th moment of position:
		\begin{equation}
				\EA{\Abs{X\of{t}}^{\gamma}}
			=
				\Int{\Reals}{}{x} \Abs{x}^\gamma \rho\of{x;t}
			\AsymEq
				A_0 + A_{\alpha\nu} e^{-\alpha \nu t} + A_{\mu\nu} e^{-\mu \nu t}
		\end{equation}
		for large times $t$.
		We take $\gamma < \alpha < \mu$, so that the moment exists for all times, as discussed before.
		The $\gamma$-dependence is hidden in the coefficients $A_0$, $A_{\alpha\nu}$ $A_{\mu\nu}$, which are defined by the corresponding integrals with respect to the eigenfunctions $\rho_0$ (the stationary value), $\rho_{\alpha\nu}$ and $\rho_{\mu\nu}$; compare with Eq.\eqref{eq:ObsRelax}.
		For broad initial states we have $\lambda_\text{min} = \alpha \nu < \mu\nu$.
		We stress here that although we illustrate the procedure in some moment, the expansion holds for all observables!

		When the stable index of the noise, $\mu$, is known, this quite universal relaxation pattern allows for inferring the preparation state from the relaxation spectrum.
		The reason is, that broad initial state results in the existence of a relaxation rate $\alpha \nu < \mu \nu$, which in turn dominates the long time behavior.
		Hence, given the time evolution of \textit{any} observable, we set out to find the smallest relaxation rate $\lambda_\text{min}$.
		A value of $\lambda_\text{min}$ significantly smaller than $\mu\nu$ proves the existence of a ``broad'' preparation state.
		If it is possible to access more than just the first relaxation rate, another possible test would be to examine the ratio $\lambda_\text{next} / \lambda_\text{min}$.
		This method is however not very robust, since $\alpha$ could also be a rational fraction of $\mu$.
		We show that determining $\lambda_\text{min}$ is indeed possible by using synthetic data from computer simulations and present some possible algorithms to infer the slowest relaxation rate.
		We present two naive approaches: the first based on linear regression, the second based on complexification and Fourier analysis.
		In case they fail, the method can be augmented by wavelet analysis.
		The procedure is very similar to the one in \cite{Dybiec2015}, where the authors sought for the smallest relaxation rate in a fractional escape problem.

		We performed Monte-Carlo simulations of an ordinary OUP's It\^o-Langevin equation (with corresponding FPE given by Eq.\eqref{eq:FPE}) with different $\mu$ and $\alpha$.
		The chosen values can be inspected in table \ref{tab:Result}.
		We used $\gamma = \tfrac{4}{10} \Min{\alpha}{\mu}$.
		This way the $\gamma$-th moment will always exist and the corresponding sample average will have a finite variance (because $2\gamma < \alpha$ and $2\gamma < \mu$).
		By the central limit theorem, the sample average's fluctuations are asymptotically Gaussian, despite the L\'evy initial state.
		We also analyzed a data set with Gaussian initial distribution and standard deviation equal to $0.1$, i.e. $\alpha = 2$.
		In that case, we chose $\gamma = 2$.
		We averaged over $8192$ trajectories with $32768$ time steps and a total length of $T = 32\nu^{-1}$.
		At last, we chose natural units, i.e. $\nu = K = 1$.

		\subsection{Finding the equilibrium value}
			\begin{figure}%
				\includegraphics[width=0.8\columnwidth]{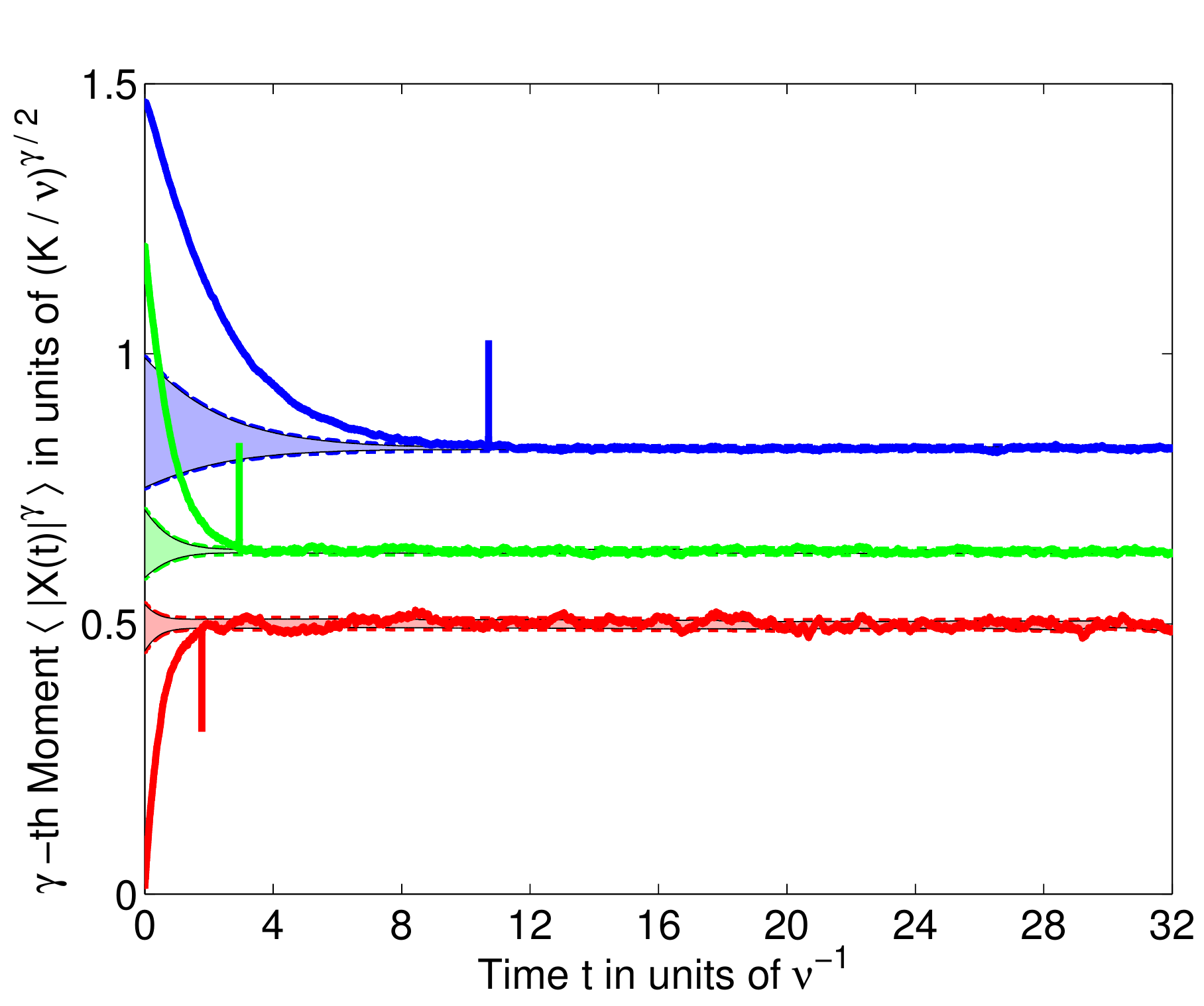}%
				\caption{
					Signal and the time averaged zone.
					The $\gamma$-th moment is plotted against time for different values of $\alpha$ and for $\mu = 2.0$.
					We take $\gamma = 0.4\Min{\alpha}{\mu}$.
					From top to bottom: (blue) $\alpha=0.5$, (green) $\alpha=1.5$, (red) $\alpha=2.0$.
					The colored area corresponds to the time average plus its standard deviation.
					The equilibration time $t_\text{eq}$ is determined as the time when the solid curve first enters the colored area, as indicated by the arrows.
					\label{fig:Moment}
				}
			\end{figure}

			Given some observable, we must first find the equilibrium value $A_0$.
			This is very simple and it is done by time averaging the data from the end to the beginning and finding the time $t_\text{eq}$ when the signal enters the strip of average value $\pm$ standard deviation.
			The time average is an estimate to $A_0$:
			\begin{equation*}
					A_0
				=	
					\frac{1}{T-t_\text{eq}} \Int{t_\text{eq}}{T}{t} \EA{\Abs{X\of{t}}^\gamma}
				.
			\end{equation*}
			The moments are plotted in Fig.\ref{fig:Moment}, together with the area of time average and standard deviation.
			The equilibration time is indicated as an arrow in the figure.

			For further processing we restrict the data to the interval $[0,t_\text{eq}]$ and may now define 
			\begin{equation}
					Z\of{t}
				=
					\log\of{\Abs{ F\of{t} - A_0 }}
				\AsymEq
					\log\of{\Abs{A_\lambda}} - \lambda t
				.
			\end{equation}
			In this representation we can neglect the error coming from the next exponential term, because
			\begin{align*}
					\log\of{ A_\lambda e^{-\lambda t} + A_{\lambda'} e^{-\lambda' t} }
				= &
					Z\of{t}  
					+ \log\of{ 1 + \frac{A_{\lambda'}}{A_\lambda}  e^{-\br{\lambda'-\lambda} t} }
				\\ \approx &
					Z\of{t}
					+ \frac{A_{\lambda'}}{A_\lambda}  e^{-\br{\lambda'-\lambda} t} 
				,
			\end{align*}
			and the remaining exponential is small for large enough times, since $\lambda' - \lambda > 0$.
			We will refer to $Z\of{t}$ simply as the signal.

		\subsection{Asymptotic Fitting}
			\begin{figure}
				\includegraphics[width=0.8\columnwidth]{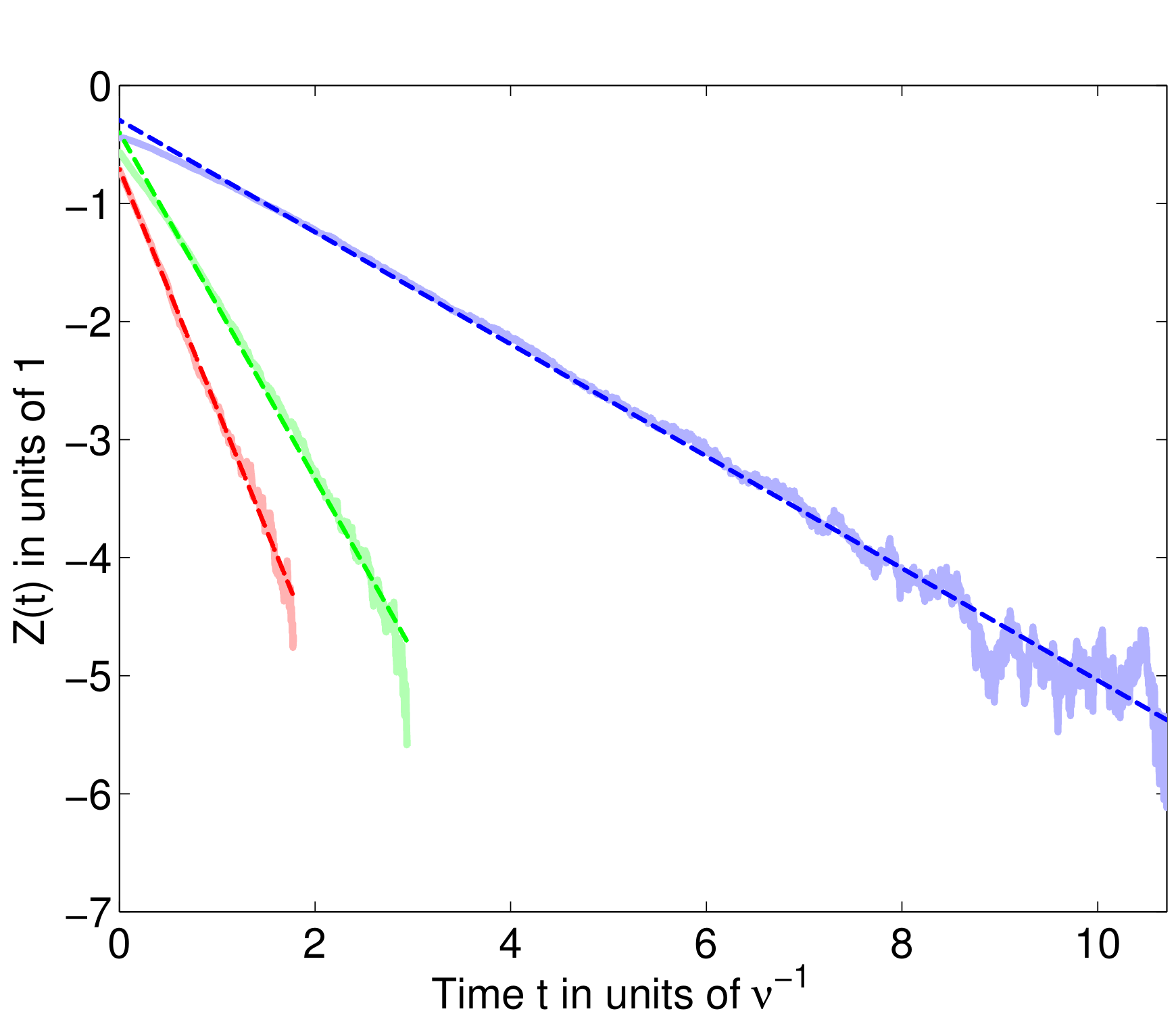}
				\caption{
					Logarithmic signal with linear regression curves.
					The logarithmic signal $Z\of{t}$ is plotted over time; the dotted lines are the linear regression curves.
					The slopes are given in Table \ref{tab:Result}, $\mu$ is $2.0$, $\alpha$ values from top to bottom: $0.5$ (blue), $1.5$ (green) and $2.0$ (red).
					Signals are only plotted up to $t_\text{eq}$, such that $Z\of{t}$ remains finite.
					The linear fit gives good results in the whole time domain.
					\label{fulinfit}
				}
			\end{figure}

			In terms of $Z\of{t}$, the exponential fit becomes a simple linear regression.
			In figures \ref{fulinfit}, we show the signal $Z\of{t}$ and the exponential fit in a log-linear plot, which demonstrates that $\log Z\of{t}$ relaxes monotonically and almost linearly from its initial value to the stationary value $A_0$. 
			Thus, one option to estimate $\lambda_\text{min}$, is a simple linear regression of $Z\of{t}$.
			The corresponding values of $\alpha$ derived by this method are listed in Table \ref{tab:Result}.

			Utilizing another approach proposed in \cite{Postnikov2015}, we can map the real valued $Z\of{t}$ onto the complex function 
			\begin{equation}
					\zeta\of{t} 
				= 
					\exp\of{ i \xi Z\of{t} } 
				\AsymEq
					\exp\of{ i \xi \log \Abs{A_\lambda} - i \xi \lambda t }
				,
				\label{eq:ComplexSignal}
			\end{equation}
			and we can apply the naive Fourier transform because the function (\ref{eq:ComplexSignal}) has constant unit amplitude and the original function multiplied by the appropriate constant factor $\xi$ has a sense of the phase. Here $\xi \approx 20\pi / \Abs{Z\of{t_\text{eq}} - Z\of{0}}$ is chosen in such a way that there are multiple oscillation over the interval $[Z\of{0}, Z\of{t_\text{eq}}]$.
			A purely periodic signal should result in a sharp peak in the Fourier transform, and one can find its maximum as an estimate for $\lambda_\text{min}$, see Table \ref{tab:Result}. 

			However, both these procedures have the same weakness: they do not measure the first relaxation rate, but rather some average, similar to \cite{Dybiec2015}.
			Therefore we must use a local technique, like wavelet analysis.

			\begin{table}
				\setlength{\tabcolsep}{2ex}
				\begin{tabular}{c|c||c|c|c}
					$\mu$	&	$\alpha$	&	$\lambda_\text{LR}$	&	$r$			&	$\lambda_\text{F}$	\\	\hline
					$1.0$	&	$0.5$		&	$0.501$				&	$-0.9934$	&	$0.352$				\\
					$1.5$	&	$0.5$		&	$0.526$				&	$-0.9950$	&	$0.493$				\\
					$1.5$	&	$1.0$		&	$0.892$				&	$-0.9932$	&	$0.835$				\\
					$2.0$	&	$0.5$		&	$0.475$				&	$-0.9967$	&	$0.462$				\\
					$2.0$	&	$1.5$		&	$1.466$				&	$-0.9963$	&	$1.401$				\\
					$2.0$	&	$2.0$		&	$2.033$				&	$-0.9976$	&	$2.083$				
				\end{tabular}
				\caption{
					Results of the naive approaches.
					The third column lists the result of a simple linear regression of $Z\of{t}$.
					$r$ denotes Pearson's correlation coefficient of the data.
					The last column is computed from the maximal position of the Fourier transform of $\zeta\of{t}$.
					\label{tab:Result}
				}
			\end{table}

	\section{Wavelet-based analysis}
		In contrast to the situations studied in Ref.~\cite{Postnikov2015}, the time dependence of the complexified signal's phase is not linear in a general case. 
		Thus, we need to generalize Fourier transform to a transform providing local spectral analysis, e.g. the wavelet transform with the Morlet wavelet:
		\begin{equation}
				w\of{a;t}
			=
				\frac{1}{\sqrt{2\pi a^2}}
				\Int{\Reals}{}{t'} \zeta\of{t'} e^{i\omega_0\frac{(t'-t)}{a}}
				e^{-\frac{(t'-t)^2}{2a^2}}
			.
			\label{wvlM}
		\end{equation}
		Here $\omega_0$ is called the central frequency; the choice $\omega_0=2\pi$ enables us to interpret the scale parameter $a$ as the period of the wavelet.

		By direct calculation, it can easily be shown that the complexified signal from Eq.\eqref{eq:ComplexSignal} results in the transform:
		\begin{equation*}
				w\of{a;t}
			=
				\zeta\of{t} 
				e^{-\frac{(\xi \lambda a - \omega_0)^2}{2}}
			,
		\end{equation*}
		which implies that the maximum of its absolute value $\Abs{w\of{\tilde{a}\of{t};t}} = \mathrm{max} \Abs{w\of{a;t}}$ allows the determination of the local relaxation rate:
		\begin{equation}
				\lambda\of{t} 
			= 
				\frac{\omega_0}{\xi \tilde{a}\of{t}}
			.
			\label{eq:LocLambda}
		\end{equation}

		In practical realizations, the transform \eqref{wvlM} can be easily evaluated by using the convolution theorem.
		The version, which operates with a discrete sample, reads as 
		\begin{equation}
				w\of{a_i;t_j}
			=
				\widehat{F}^{-1}\off{
					\widehat{F}\off{\zeta}\of{\omega_l}
					e^{-\frac{(\omega_l a_i - \omega_0)^2}{2}}
				}\of{t_j}
			,
		\label{wvlMd}
		\end{equation}
		where $\widehat{F}$ and $\widehat{F}^{-1}$ denote the direct and inverse fast Fourier transforms, respectively.

\begin{figure}
			\includegraphics[width=\columnwidth]{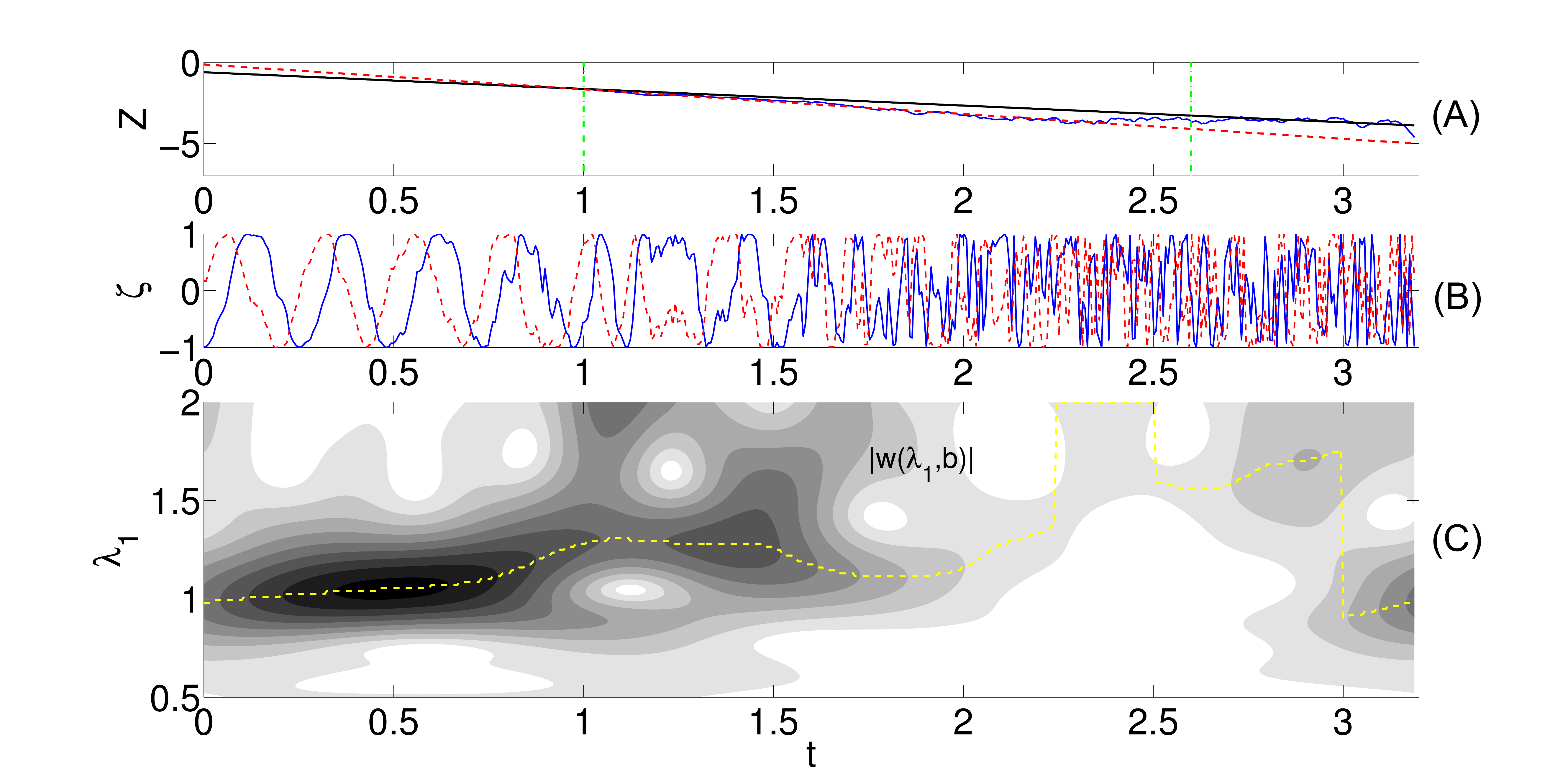}
			\caption{
				Logarithmic and complexified signal, and wavelet transform.
				All data given for $\mu=2.0$ and $\alpha = 1.5$.
				(A): Logarithmic signal. 
				Vertical dash-dotted lines mark the approximate transition time points. 
				The solid line (black) corresponds to the linear fit in $t\in [0,1]$, the dashed line (red) is the linear fit in $t\in [1,2.6]$.
				(B): Complexified signal in real (blue, full line) and imaginary (red, dashed line) part.
				(C): Wavelet transform's absolute value.
				It is calculated from the complexified signal above.
				Darker regions correspond to larger absolute value.
				The dashed curve (yellow) traces the global absolute value maximum for each moment.
				\label{fugloguwvl}
			}
		\end{figure}

		As an example, we consider the complex function $\zeta\of{t}$ for $\mu=2.0$ and $\alpha=1.5$.
		Fig.~\ref{fugloguwvl}A shows $Z\of{t}$ and Fig.~\ref{fugloguwvl}B shows $\zeta\of{t}$ with $\xi = 8\pi$ for that case.
		Both are plotted until $t_\text{eq}$ to avoid taking logarithms of negative numbers.
		One can see that the dynamics of $\zeta\of{t}$ changes from the regular oscillations with a growing period to extremely slow dynamics with random phase changes.
		The latter correspond to the fluctuations of the observable as it enters the stationary state.
		In principle, the correct boundary of the emerging stationary state could be estimated even from this picture.

		Fig.~\ref{fugloguwvl}C shows the absolute value of the continuous wavelet transform applied to the equidistant sample of the analyzed function via procedure \eqref{wvlMd}.	
		The transform is plotted as an explicit function of the relaxation rate by using Eq.\eqref{eq:LocLambda}.
		We can clearly separate three subintervals with different absolute value maxima trends and find the corresponding relaxation rates.
		The first interval starts from zero and continues to $t\approx 1$.
		Following the absolute values' maximum, we obtain a rate with a slightly growing time dependence.
		Around $t \approx 1$ the absolute value's maximum transits to the one of the first non-spectral relaxation rate.
		This part goes on to approximately $t \approx 2.6$.
		It should be pointed out that the definite irregularity of the instant relaxation rate reflects a sensitivity of the continuous wavelet transform to short-time fluctuations of the observable.
		It originates from the Gaussian window in Eq.~\ref{wvlM}, the width of which is adjusted to the detected periods: about five individual oscillations fit into the bell-shaped window.
		The rest of the interval, where there are no intensive absolute values maxima, presents fluctuations around the equilibrium state. 

		The revealed transition points are used to determine the intervals for the least mean square fits within the boundaries, which they define.
		Such linear fit over $t\in[0,1]$ is shown as the solid line with the slope $-\lambda_{[0,1]}=-1.06$.
		Its absolute value coincides (with a quite reasonable accuracy) with the parameter $\nu$ in the studied Ornstein-Uhlenbeck process.
		Therefore, we can conclude that this first regime of the relaxation process is completely classical (spectral) one.
		The fit line practically undistinguished from the  relaxation dependence up to $t=1$ (the correlation coefficient: 0.9996) but further its inadequacy is quite visible.
		This bounding value  also supports the explanation as a spectral relaxation since $t=\nu^{-1}$ is the classical relaxation time for the Ornstein-Uhlenbeck process.
		However, since it appears in the short-time limit, it is not covered in above theory.
		Our theory is only concerned about the long-time behavior.

		At the same time, the dependence of $Z\of{t}$ for $t>1$ remains linear as well (the correlation coefficient: $0.9967$), but with another slope coefficient.
		The linear fit within $t\in[1,2.6]$ provides the slope value $\lambda_\text{min}=1.30$, i.e. its absolute value is sufficiently close to the leading relaxation rate of the non-spectral mode $1.5$, which, as one can see, prevails within this region. 
		Moreover, comparison of the relaxation dependence and the last linear fit (dashed straight line) in Fig.~\ref{fugloguwvl}A demonstrates that the later relaxation process follows this non-spectral character: the logarithmic observable only trembles around this linear fit (although with larger, almost symmetric, deviations). 

		In conclusion, Linear fitting procedures can be suitable for the determination of $\lambda_\text{min}$.
		In addition, the wavelet scale parameter regression of the logarithm of non-stationary excess part of the observable is preferable if one needs to study in details the transient process between two regimes such as one located within the time interval $t\in[0.6,1]$.
		
		Here, we analyzed $\lambda\of{t}$ just with the bare eyes.
		A more quantitative analysis is possible using techniques of change-point detection, \cite{Chen2013,Brodskij2000}.

	\section{Summary, discussion and conclusion}
		We discussed the L\'evy-Ornstein-Uhlenbeck process with respect to its non-trivial property stating that all non-negative relaxation rates are admissible, i.e. for broad initial state, such with power-law tails, ``non-spectral'' relaxation rates occur that do not belong to the spectrum of an associated Schr\"odinger operator.
		These rates are visible in the relaxation pattern of every observable and can be inferred from the pattern.
		Hence, given some data $F\of{t}$, it is possible to test for broad initial state.

		The proposed technique of the Ornstein-Uhlenbeck random process analysis can be summarizes as follows:
		\begin{itemize}
			\item
				First find the equilibration time $t_\text{eq}$, when $F\of{t}$ assumes its equilibrium, by computing time averages from the end of the data set.
				$t_\text{eq}$ is the first time, when $\Abs{F\of{t} - F_\text{eq}} < \delta F_\text{eq}$. 
			\item 
				Construct the logarithmic signal $Z\of{t} = \log\Abs{F\of{t} - F_\text{eq}}$, the complexified signal $\zeta\of{t} = \exp\of{i \xi Z\of{t}}$ and calculate its continuous Morlet-wavelet transform $w\of{a;t}$.
			\item 
				The relaxation rates can be inferred from a linear regression of $Z\of{t}$, from the maximum of $\zeta\of{t}$ Fourier spectrum or locally from the maximal line of the wavelet transform's absolute value.
				If the absolute value of this transform contains points, where sharp transitions occur, then this means that different (spectral and non-spectral) relaxation processes exist.
				The transitions mark boundaries for those regimes.
				The relaxation curve for each subinterval can be fitted separately.
		\end{itemize}

		Note, that we have been concerned with the long-time behavior of relaxation.
		It is not surprising that the most visible manifestation of non-spectral relaxation occurs after the standard relaxation time $\nu^{-1}$ only.

	\section*{Acknowledgment}
		This work is supported by DFG (project SO 307/4-1).

	\appendix
	\section{Properties of the fractional Fokker-Planck operator}
		Although a general approach to the relaxation properties of the FFPE (based on the convergence of the Kullback-Leibler distance) seems possible for the generalized OUP as well, it is not necessary in our case.
		All relevant properties can be shown explicitly, since the solution to Eq.\eqref{eq:FPE} is known in Fourier domain.
		We already have shown in the main text that the stationary state of Eq.\eqref{eq:FPEFourier} is given by:
		\begin{equation}
			\hat{\rho}_\text{eq}\of{k}
			=
				e^{-\frac{K}{\mu\nu} \Abs{k}^\mu }
			,
		\end{equation}
		furthermore it was shown in \cite{Toenjes2013}, Eq.(11), that the initial value problem is solved by:
		\begin{equation*}
				\hat{\rho}\of{k;t}
			=
				\frac{
					\hat{\rho}_\text{in}\of{ke^{-\nu t}}
				}{
					\hat{\rho}_\text{eq}\of{ke^{-\nu t}}
				}
				\hat{\rho}_\text{eq}\of{k}
			=
				\hat{\rho}_\text{in}\of{ke^{-\nu t}}
				e^{-\Abs{k}^\mu \br{ 1 - e^{-\mu\nu t}} }
			.
			\label{eq:Sol}
		\end{equation*}
		By taking the limit $t\to\infty$, we immediately see that the solution converges to the stationary state.
		The normalization -- which is obtained in Fourier domain by taking $k\to0$ -- also is conserved.
		In real space, the solution as given by Eq.\eqref{eq:Sol} is a convolution of the initial state with the pdf of a $\mu$-stable random variable.
		Hence, if the initial state is not oscillating, i.e. is non-negative, it will remain like this forever.
		This shows all statements of the main text.

	\bibliographystyle{aipnum4-1}
	\bibliography{Article,Book,Self,NotRead}
\end{document}